\documentstyle[aps,prl,multicol,epsfig]{revtex}

\begin{document}

\newcommand \qea {\mbox{$q_{\scriptscriptstyle {\rm EA}}$}}

\title{Two time scales and FDT violation in a Finite Dimensional Model
for Structural Glasses}

\author{Federico Ricci-Tersenghi$^1$, Daniel A. Stariolo$^2$ and
Jeferson J. Arenzon$^2$}

\address{
$^1$Abdus Salam International Center for Theoretical Physics,
Condensed Matter Group\\
Strada Costiera 11, P.O. Box 586, 34100 Trieste, Italy\\
$^2$Instituto de F{\'\i}sica, Universidade Federal do Rio Grande do Sul\\ 
CP 15051, 91501-970 Porto Alegre RS, Brazil\\
E-mails: {\tt riccife@ictp.trieste.it, stariolo@if.ufrgs.br,
arenzon@if.ufrgs.br}}

\date{\today}

\maketitle

\begin{abstract}
We study the breakdown of fluctuation-dissipation relations between
time dependent density-density correlations and associated responses
following a quench in chemical potential in the Frustrated Ising
Lattice Gas. The corresponding slow dynamics is characterized by two
well separated time scales which are characterized by a constant value
of the fluctuation-dissipation ratio. This result is particularly
relevant taking into account that activated processes dominate the
long time dynamics of the system.
\end{abstract}

\pacs{PACS numbers: 75.10.Nr, 05.50.+q, 75.40.Gb, 75.40.Mg}
 
\begin{multicols}{2}
\narrowtext

In recent years considerable progress has been achieved in the
theoretical description of the glassy state of matter. A scenario for
the observed slow dynamics of glass forming materials has emerged
through detailed analysis of mean field (MF) spin glass models
~\cite{young}. The equations describing the off-equilibrium dynamics
of these MF spin glasses simplify, above the transition, to the single
equation for the Mode Coupling Theory for supercooled
liquids~\cite{gotze}. These approaches have been successful in
explaining history dependence or aging effects and the nature of the
two characteristic relaxations in glasses, the short time or
$\beta$-relaxation and the structural long time
$\alpha$-relaxation. In the $\alpha$-relaxation the system falls out
of local equilibrium, as a consequence Fluctuation-Dissipation Theorem
(FDT) breaks down and can be replaced by the more general relation
\begin{equation}
R(t,t_w) = \frac{X(t,t_w)}{T}\frac{\partial C(t,t_w)}{\partial t_w}\ ,
\label{gfdt}
\end{equation}
where $C(t,t_w)$ is a two times correlation function and $R(t,t_w)$
with $t\!>\!t_w$ is the associated response. $T$ is the heat bath
temperature and $X(t,t_w)$ is a function that measures the departure
from FDT: at equilibrium $X=1$ and the usual FDT is recovered, while
in the out of equilibrium regime $X<1$. In MF
approximation~\cite{cuku} the function $X$, called
``fluctuation-dissipation ratio''(FDR), turns out to depend on both
times only through $C(t,t_w)$. Moreover in MF models of
glasses $X$ is a constant (when different from 1).  This scenario
reflects the existence of only two well separated time scales, the
equilibrium or FDT scale and a longer one where the system is out of
equilibrium. The FDR has been interpreted as an effective temperature
and it has been demonstrated that it is exactly the temperature that a
thermometer would measure if it would be coupled to the slowly
relaxing modes of the system~\cite{luca}. Recently the first
experimental determination of the FDR has been done in
glycerol~\cite{grigera}.

This scenario, while appealing, is essentially based on an analogy
between the physics of some MF spin glasses and the behavior
of real finite structural glasses (being the formal connection valid
only in the high temperature region). At this point it seems crucial to
test the link between MF theories and realistic models in the
glassy phase. In particular, we still do not know which will be the role
of activated processes in realistic models. Activated processes are
absent in completely connected models in the thermodynamic limit while
on the contrary they dominate the relaxation dynamics below the glass
transition temperature $T_g$ in real glasses.

The goal of the present letter is to go beyond the MF like
description of structural glasses by considering a finite dimensional
model, the frustrated Ising lattice gas (FILG)~\cite{mario,nos}, which
presents most of the relevant features of glass forming materials, in
particular activated processes at low temperatures.  Here we analyze
the violation of FDT in the FILG in three dimensions through Monte
Carlo simulations. The (very precise) results confirm the qualitative
scenario of MF models of a constant FDR with large separation
of time scales and set the stage for a detailed investigation of
activated processes in realistic models of glasses.

The FILG is defined by the Hamiltonian:
\begin{equation}
H = -J \sum_{<ij>} (\varepsilon_{ij} \sigma_i \sigma_j - 1) n_i n_j -
\mu \sum_i n_i\ .
\end{equation}
At each site of the lattice there are two different dynamical
variables: local density (occupation) variables $n_i=0,1$ ($i=1 \ldots
N$) and internal degrees of freedom, $\sigma_i=\pm1$. The usually
complex spatial structure of the molecules of glass forming liquids,
which can assume several spatial orientations, is in part responsible
for the geometric constraints on their mobility. Here we are in the
simplest case of two possible orientations, and the steric effects
imposed on a particle by its neighbors are felt as restrictions on its
orientation due to the quenched random variables $\varepsilon_{ij}=\pm
1$. The first term of the Hamiltonian implies that when $J\rightarrow
\infty$ any frustrated loop in the lattice will have at least one hole
and then the density will be $\rho<1$, preventing the system from
reaching the close packed configuration. The system will then present
``geometric frustration''.  Finally, $\mu$ represents a chemical
potential ruling the system density (at fixed volume).

The system presents a slow ``aging'' dynamics after a quench from a
small value of $\mu$ characteristic of the liquid phase to a large
$\mu$ corresponding to the glassy phase~\cite{nos}, what is equivalent
to a sudden compression.
In the present numerical experiments we always let the system evolve
after a quench in $\mu$ with $J$ and $T$ fixed.  The origin of the
times is set on the quench time.  After the quench the density slowly relaxes
up to a critical value near $\rho_c \approx 0.675$ (further
details will be given in~\cite{nos3}).  After a waiting time $t_w$ we
fix the density to the value $\rho=\rho(t_w)$ (for technical reasons
explained below) and a small random perturbation ($\mu_i=\pm1$) is
applied~\cite{note1}:
\begin{equation}
H'(t) = H(t) - \epsilon(t) \sum_i \mu_i\, n_i(t)\ .
\label{pert_ham}
\end{equation}
In all our numerical experiments the field is switched on at time
$t_w$ and kept fixed for later times, that is $\epsilon(t) =
\epsilon\, \theta(t-t_w)$.  Then we measure the density-density
autocorrelation function
\begin{equation}
C(t,t_w) = \frac{1}{N\rho}\sum_i \overline{\langle n_i(t) n_i(t_w)
\rangle}\ ,
\label{corr}
\end{equation}
where $\langle\ \cdot\ \rangle$ and $\overline{\ \cdot\ }$ are the
averages over thermal histories and disorder
realizations~\cite{note2}.  At the same time we measure the associated
response function integrated over the time and divided by the
perturbing field intensity, which defines the off-equilibrium
compressibility
\begin{equation}
\kappa(t,t_w) = \frac1\epsilon \int_{-\infty}^{t} R(t,s) \epsilon(s)
{\mathrm d}s = \int_{t_w}^{t} R(t,s) {\mathrm d}s\ ,
\label{compre1}
\end{equation}
where, as usual, the response is defined as
\begin{equation}
R(t,t') = \frac{1}{N\rho} \sum_i \frac{\partial \overline{\langle
n_i(t) \rangle}}{\partial \epsilon(t')}\ .
\label{response}
\end{equation}
In the large times limit ($t,t_w\to\infty$), $X(t,t_w)$ depends
on both times only through the correlation $C(t,t_w)$.
Then integrating Eq.(\ref{gfdt}) from $t_w$ to $t$ we obtain a useful
relation linking the correlation and the compressibility in the out of
equilibrium regime
\begin{equation}
T \kappa(t,t_w) = \int_{C(t,t_w)}^1 X(C) {\mathrm d}C\ .
\label{main}
\end{equation}
This is the key relation used in order to extract the FDR.

In our case the perturbing term in the Hamiltonian, shown in
Eq.(\ref{pert_ham}), gives to the integrated response the following form
\begin{equation}
\kappa(t,t_w) = \frac{1}{N\rho} \left( \sum_i \overline{[\langle \mu_i
n_i(t) \rangle]_{\rm av}} - \sum_i \overline{[\langle \mu_i n_i(t_w)
\rangle]_{\rm av}} \right)\ ,
\label{compre2}
\end{equation}
where $[\ \cdot\ ]_{\rm av}$ is the average over the random $\mu_i$
realizations~\cite{note2}.  The second term can be ignored because the
$\mu_i$ are random and completely uncorrelated from the configuration
at time $t_w$.

Performing a parametric plot of the compressibility (or the integrated
response) versus the correlation is a useful way of getting
information about the different dynamical regimes present in the model
and in particular the time scales structure of the system.  In fact
from Eq.(\ref{main}) it is easy to see that, plotting $T
\kappa(t,t_w)$ vs. $C(t,t_w)$, the FDR can be simply obtained as minus
the derivative of the curve, i.e.
\begin{equation}
X(C') = -\left.{\partial[T\kappa(t,t_w)]}\over{\partial C(t,t_w)}
\right|_{C(t,t_w)=C'}\ .
\end{equation}
There is already a considerable literature on this kind of analysis in
systems with and without quench disorder~\cite{fdt_varie}.
Particularly relevant to the present discussion are
references~\cite{fdt_vetri}, where a constant FDR was found
in model glasses with interactions of the Lennard-Jones type and in a
purely kinetic lattice gas. The FILG has the advantage of being a
Hamiltonian lattice model with short range interactions and, in this
sense, it is more realistic than purely kinetic models and more
accessible analytically and computationally than Lennard-Jones
systems.  Moreover, it is a valid on-lattice model for structural
glasses and it may be simple enough to apply statistical mechanics
techniques. 

We have simulated the FILG in 3D for linear sizes $L=30$ and
$60$. Fixing the coupling constant $J=1$ and a temperature $T=0.1$
the system presents a glass transition around $\mu=0.5$~\cite{mario}. 

In all our numerical experiments we have prepared the system in an
initial state with low density, characteristic of the liquid phase and
then, at time zero, we have performed a sudden quench in $\mu$ to a
value deep in the glass phase (the data presented here refer to
$\mu=1$).  As already explained above, after a time $t_w$ a
perturbation in the form of a random small chemical potential has been
applied and the density-density autocorrelation, Eq.(\ref{corr}), and
the corresponding integrated response, Eq.(\ref{compre2}), have been
recorded.  At time $t_w$ the density has been fixed to the value
$\rho=\rho(t_w)$ for the following reason: the perturbing term in the
Hamiltonian [see Eq.(\ref{pert_ham})] favors roughly half of the sites
(those with $\mu_i=1$) and the remaining half to be empty.  When the
perturbation is switched on, the first half starts to be more filled
than the second one and this increases the integrated response, as it
should.  On very long times, however, because the density reached at
time $t_w$ is still a bit below the asymptotic value, the number of
particles continues growing and new particles are added with higher
probability on the sites with $\mu_i=-1$, which are more empty.  Then
the response becomes negative because of systematic errors.  We avoid
the negative responses fixing the density at time $t_w$.  In the
limit $t_w\!\to\!\infty$ we recover the right behavior in any case,
however with our choice the extrapolation is safer.

In all our simulations we have verified that the strength of the
perturbation $\epsilon=0.02$ is small enough in order to be in the
linear response regime. We have also checked that the system
thermalizes at temperatures $T \ge 0.3$ (being always $\mu=1$) and
satisfies FDT.  Further details will be given in~\cite{nos3}.

\begin{figure}
\centerline{\epsfig{file=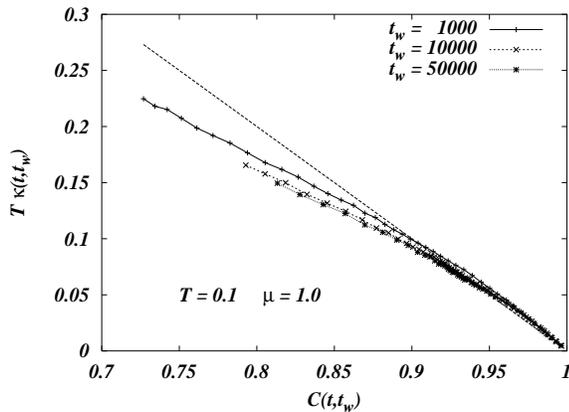,width=8cm}}
\caption{The plot of the integrated response times the temperature
versus the correlation gives clear evidence for the existence of two
well separated time scales, together with a constant FDR in both
regimes. The dashed line is $T\kappa=1-C$. The errors are of the order
of the symbols and have been estimated from sample to sample
fluctuations.}
\label{fig1}
\end{figure}       

In Fig.~\ref{fig1} we show the main result of this letter: a
parametric plot of the integrated response versus the density
autocorrelation for different waiting times. The behavior of the
curves is exactly the one predicted by MF theories for a glass former.
Two distinct regimes can be perfectly recognized.  In the FDT or
quasi-equilibrium regime, $t-t_w \ll t_w$ ($\beta$-relaxation), the
points lie on the straight line given by
\begin{equation}
T \kappa(t,t_w) = 1 - C(t,t_w)\ .
\end{equation}
This first time regime corresponds then to a FDR $X(t,t_w)=1$
independent of $t$ and $t_w$. In this regime the system is in
quasi-equilibrium, with the particles moving inside the cages formed
by nearly frozen neighbors and the temperature measured from
particles fluctuations is that of the heat bath. When $t-t_w \ge t_w$
the system falls out of equilibrium, entering the aging regime, and
the data in Fig.~\ref{fig1} depart from the FDT line.  From the figure
it is clear that the FDR is still a constant, but now $0 < X(t,t_w) <
1$. The constancy of the FDR in the out of equilibrium regime is one
of the central predictions of MF approaches. Here we see that
this is still valid for a finite dimensional system.

\begin{figure}
\centerline{\epsfig{file=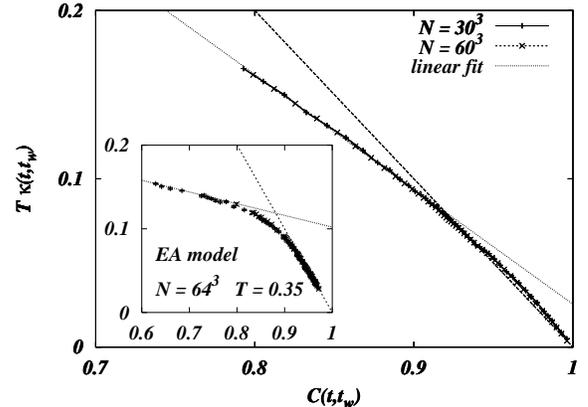,width=8cm}}
\caption{Same as Fig.~\ref{fig1}, now with different volumes in order
to show the absence of finite size effects. The linear fit to the data
is very good and gives a value for the FDR $X \simeq 0.65$. For the EA
model (inset) the linear fit is far from good.}
\label{fig2}
\end{figure}       

In Fig.~\ref{fig2} we compare the results for the FDR measured on two
large systems, whose sizes are $N=30^3$ and $N=60^3$.  Both curves
correspond to $t_w=10^4$ and no finite size effects are evident.
Fitting the data in the out of equilibrium regime to the straight line
\begin{equation}
T \kappa(t,t_w) = X(t_w) [\qea(t_w)-C(t,t_w)] + [1-\qea(t_w)]\ ,
\label{fit}
\end{equation}
allows us to compute the $t_w$-dependent FDR $X(t_w)$ and the
Edwards-Anderson order parameter $\qea(t_w)$.  In the large times
limit they should converge to the corresponding equilibrium 
values~\cite{fmpp,mpv}. 

%$x$
%and \qea~\cite{fmpp}, giving information on the pure states
%structure~\cite{mpv}: $1-\qea$ measures the pure state width or
%equivalently the size of connected regions in configurational space
%and $x$ is the probability of two configurations being in the same
%state.

We report the linear fit in Fig.~\ref{fig2} in order to show how well
the data can be fitted with the formula in Eq.(\ref{fit}).  As one can
see from Fig.~\ref{fig1}, the slope $X(t_w)$ changes very little with
$t_w$ and it takes the same value (within the error) for the two
largest waiting times.  The results of our fits give $X=0.64(3)$ and
$\qea=0.92(1)$.  In comparison with other works the correlation range
we are exploring may seem quite small.  However it should be kept in
mind that we are using non-connected correlations functions which tend
in the large times limit to $\rho^2 \simeq 0.44$.  So we are actually
spanning half of the allowed range.

In the inset of Fig.~\ref{fig2} we present the same kind of data
($t_w\!=\!10^5,10^6$) for the Edwards-Anderson (EA) model, which is
expected to have more than two time scales. It is clear that in the EA
model the slope changes along the curve and a straight line is not
able to fit the whole set of data. Moreover higher temperatures
data~\cite{fdt_varie} suggest that the slope still have to decrease
(in modulus) for smaller correlations, making the linear fit even
poorer.

An important difference between the present model and MF approaches
should be evident: activated processes present in finite-dimensional
systems should dominate the asymptotic dynamics and as $t_w\to\infty$
equilibrium dynamics should be restored (a similar behavior can be
observed in MF models by considering small systems~\cite{felix}).
However the time a macroscopic system needs to reach such an
equilibrium (thermalization time, $t_{\rm eq}$) increases very rapidly
with the system size and then the use of large sizes (here we have
$L=30$ and $60$) prevents the system from reaching equilibrium in
accessible time scales.  In other words, we expect that the FDR
explicitly depends on $t_w$ and it tends to 1 in the limit
$t_w\to\infty$.  However in the range $1 \ll t_w \ll t_{\rm eq}$
(where we actually are) the FDR should relax into some very long
plateau. It is very remarkable that finite times effects are very well
described by MF theories.

\begin{figure}
\centerline{\epsfig{file=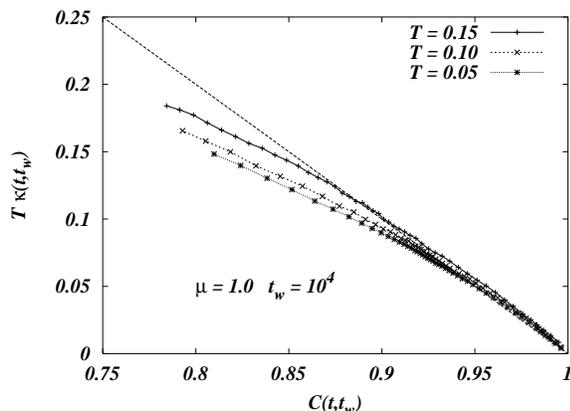,width=8cm}}
\caption{Same as Fig.~\ref{fig1} for different temperatures in the
glassy phase.}
\label{fig3}
\end{figure}

In Fig.~\ref{fig3} we show the usual integrated response versus
correlation plot for different temperatures, being the chemical
potential always equal to $\mu=1$.  We estimated, both from the
density measurements and from the FDR, the glassy transition to be
located around $T_g \simeq 0.2$.  So all the data refer to the glassy
phase. The main result to be noted is the good parallelism between all
the curves in the aging regime.  They can be perfectly fitted with the
same value for $X$ and different values for \qea, that increases
lowering the temperature. As $T$ approaches $T_g$ from below, 
FDT is recovered. Also, as
$\qea(T)$ decreases, $X$ remains constant within the numerical precision,
what is in contrast with MF models where the FDR is
usually proportional to the temperature ($X \propto T$). The
exact behavior near the transition, that is, whether
$X$ (and $\qea$) are continuous or not~\cite{mf}, 
is difficult to establish numerically and will be addressed in a future work.
%Maybe the
%model has a dynamical transition at $T_g$ and, for any $T < T_g$, we
%are actually measuring the properties of the threshold states.

In summary, we have found that the glassy phase of a realistic model
of structural glass presents two well separated time scales, as found
in MF models of spin glasses. The out of equilibrium long time
dynamics can be characterized by a constant value of the fluctuation
dissipation ratio $X$ which is, with a very good approximation,
independent of temperature in the glass phase. This last observation
does not agree with MF predictions. Being a model in 3D, the
frustrated lattice gas is an interesting test ground for performing a
systematic study of activated processes, a main ingredient absent in
MF models.

This work was partly supported by Brazilian agencies CNPq and FAPEMIG. JJA
acknowledges the Abdus Salam ICTP (Trieste) for support during his stay, 
where part of this work was done.

\end{multicols}

\begin{thebibliography}{20}

\bibitem{young} {\it Spin Glasses and Random Fields}, Ed. A.P. Young,
(World Scientific, Singapore, 1997).

\bibitem{gotze} W. G{\"o}tze and L. Sj{\"o}gren, Rep. Prog. Phys. {\bf
55}, 241 1992. W. G{\"o}tze in {\it Liquids, Freezing and the Glass
Transition}, Les Houches 1989, Eds. J.P. Hansen, D. Levesque and
J. Zinn-Justin (North Holland, Amsterdam, 1991).

\bibitem{cuku} L.F. Cugliandolo and J. Kurchan, Phys. Rev. Lett. {\bf
71}, 173 (1993); J. Phys. A {\bf 27}, 5749 (1994); Phil. Mag. {\bf
71}, 501 (1995).

\bibitem{luca} L.F. Cugliandolo, J. Kurchan and L. Peliti, Phys.
Rev. E {\bf 55}, 3898 (1997).

\bibitem{grigera}T.S. Grigera and N.E. Israeloff,
Phys. Rev. Lett. {\bf 83}, 5038 (1999).

\bibitem{mario} M. Nicodemi and A. Coniglio, J. Phys. A {\bf 30}, L187
(1997); Phys. Rev. E {\bf 57}, R39 (1998).

\bibitem{nos} D.A. Stariolo and J.J. Arenzon, Phys. Rev. E {\bf 59},
R4762 (1999).

\bibitem{nos3} J.J. Arenzon, F. Ricci-Tersenghi and D.A. Stariolo, in
preparation.

\bibitem{note1} For simplicity we do not make explicit the parametric
dependence of $\epsilon(t)$ on $t_w$.

\bibitem{note2} All the observables we measure are self-averaging
quantities and so every average can be neglected provided one uses
very large systems.

\bibitem{fdt_varie} S. Franz and H. Rieger, J. Stat. Phys. {\bf
79}, 749 (1995).  E. Marinari, G. Parisi, F. Ricci-Tersenghi and
J.J. Ruiz-Lorenzo, J. Phys. A {\bf 31}, 2611 (1998).  A. Barrat,
Phys. Rev. E {\bf 57}, 3629 (1998).  G. Parisi, F. Ricci-Tersenghi and
J.J. Ruiz-Lorenzo, Eur. Phys. J. B {\bf 11}, 317 (1999).

\bibitem{fdt_vetri}G. Parisi, Phys. Rev. Lett. {\bf 79}, 3660 (1997).
J.L. Barrat and W. Kob, Europhys. Lett. {\bf 46}, 637 (1999).
M. Sellitto, Eur. Phys. J. B {\bf 4}, 135 (1998). R. Di Leonardo,
L. Angelani, G. Parisi and G. Ruocco, {\tt cond-mat/0001311}.

\bibitem{fmpp} S. Franz, M. M\'ezard, G. Parisi and L. Peliti,
Phys. Rev. Lett. {\bf 81}, 1758 (1998).

\bibitem{mpv} M. M\'ezard, G. Parisi and M.A. Virasoro, {\it Spin
Glass Theory and Beyond}, World Scientific (Singapore, 1987).

\bibitem{felix} A. Crisanti and F. Ritort, {\tt cond-mat/9911226}.

\bibitem{mf} J.J. Arenzon, M. Nicodemi, and M. Sellitto,  J. de Physique
        {\bf 6}, 1143 (1996); M. Sellitto, M. Nicodemi, and J.J. Arenzon
        J. de Physique {\bf 7}, 45 (1997).
\end{thebibliography}
\end{document}